# On a cryogenic noble gas ion catcher

P. Dendooven*[1], S. Purushothaman[1], K. Gloos[1,2]

[1] KVI, Zernikelaan 25, 9747 AA Groningen, the Netherlands

[2] Department of Physics, 20014 University of Turku, Finland

* Corresponding Author:

tel.: +31 50 363 3615

fax: +31 50 363 4003

e-mail: dendooven@kvi.nl

**ABSTRACT**

In-situ purification of the gas used as stopping medium in a noble gas ion catcher by operating the device at low temperatures of 60 to 150 K was investigated. Alpha-decay recoil ions from a $^{223}$Ra source served as energetic probes. The combined ion survival and transport efficiencies for $^{219}$Rn ions saturated below about 90 K, reaching 28.7(17) % in helium, 22.1(13) % in neon, and 17.0(10) % in argon. These values may well reflect the charge exchange and stripping cross sections during the slowing down of the ions, and thus represent a fundamental upper limit for the efficiency of noble gas ion catcher devices. We suggest the cryogenic noble gas ion catcher as a technically simpler alternative to the ultra-high purity noble gas ion catcher operating at room temperature.

An energetic ion slowed down in a noble gas has a fair chance of still being in an ionic state after thermalization. Provided suitable conditions, these ions can survive long enough to be transported through an exit-hole out of the stopping volume, and subsequently be shaped into a low-energy ion beam. This method of transforming high-energy nuclear reaction products into low-energy ion beams was pioneered in the early 1980s in so-called IGISOL (Ion Guide Isotope Separator On-Line) systems, for use with light-ion fusion-evaporation reactions where recoil energies are small (<100 keV/u; for heavy ions often less than a few keV/u) [1]. It was quickly adapted to fission and heavy-ion fusion-evaporation reactions with recoil energies up to about 1 MeV/u. The evolution of the IGISOL technique up to the mid-1990s is reviewed in [2,3]. The late 1990s saw the method introduced for use at in-flight separators of radioactive ion beam facilities. The large stopping volumes of these so-called gas catchers dictated by the high energies (even after energy degrading and bunching [4,5]) require electric-field guidance of the ions for a fast and efficient extraction [6,7].

Near and at thermal energies, ions can not neutralize in collisions with noble gas atoms due to the high ionization potential of the latter. Ions will neutralize upon hitting walls or any structure inside the gas catcher. This is neglected here since the electric-field guidance in gas catchers is designed to prevent the ions from hitting the walls. What happens to thermalized ions is then determined by the presence of impurities and the ionization of the noble gas by the energetic ions and possibly by an accelerator beam and/or radioactive decay radiation [8]. Impurities take part in the neutralisation process via three-body recombination involving a free electron and form molecules/adducts with the ions, see e.g. [9]. It is important to note that the ionizing radiation also plays a role in re-ionizing those ions which have been neutralized.

In general, the nuclide of interest will appear in four different variants: in atomic and molecular forms, both in their neutral and ionized states. Neutral species can be recovered by suitable ionization mechanisms, such as irradiation with laser light [10]. Most molecules formed by ions can be broken up and the atomic ion recovered by acceleration in an electric field of about 1 kV/cm in the high-density region just outside the exit-hole of the gas cell.

Over the past 25 years, a lot of technical development has focussed on removing impurities from and preventing ionization of the noble gas. Sub-ppb impurity levels have been achieved in noble gas catchers that are built according to ultra-high vacuum standards, that are bakeable and filled with ultra-pure noble gases, see e.g. [7,9]. Constructing large ultra-pure gas catchers, although possible, is far from trivial [7]. There is, however, an alternative approach to reach ultra-pure conditions: freezing out the impurities. We describe here experiments to test this approach for the first time, and present results for slowing down and transporting energetic $^{219}$Rn ions in helium, neon, and argon gas.

Figure 1 shows the experimental setup. A $^{223}$Ra source, obtained by electrostatically collecting thermalized (in ~30 mbar of helium) alpha-decay recoil ions from a thin and open $^{227}$Ac source [11], sits at the bottom of the cell filled with a noble gas as stopping medium. The ~100 keV alpha-decay recoil ions will leave the extremely thin $^{223}$Ra source when emitted in the "upward" direction. They are then thermalized in the noble gas near the source (at about 0.5 mm in 1 bar helium at room temperature as calculated using TRIM [12]).

Electrodes provide an electric field to guide the thermalized ions towards a thin (1.7 mg/cm$^2$) aluminum foil in front of a silicon detector which records alpha particle energy spectra. Ion optics simulations using SIMION [13] show that no losses occur due to the ion transport itself (Figure 1). Ion transport is detected by the alpha decay of the ions collected on the aluminum foil (where the radon ions stick when the temperature is below about 200 K). The silicon detector also observes, with about 10 times lower efficiency, alpha particles directly from the source. Those alpha particles suffer a larger energy loss in the gas than those originating from the foil. Therefore both can be unambiguously identified. Different isotopes are identified based on their known alpha energies.

As the $^{223}$Ra activity is easily measured (960 Bq at the start and 550 Bq at the end of the experiments) and the alpha particle detection efficiency (equal to the solid angle) derived from the known geometry, the efficiency that $^{219}$Rn recoil ions survive the thermalization and transport to the aluminum foil in an ionic state is straightforwardly obtained by dividing the $^{219}$Rn count rate originating from the aluminum foil by half of the $^{223}$Ra source activity and by the detection efficiency. Decay losses during transport can be neglected because the $^{219}$Rn half-life (4.0 s) is long relative to the transport time (of order 1 ms for 1 bar of helium at room temperature).

The measured alpha particle spectra also contain peaks from $^{215}$Po and $^{211}$Bi, the other alpha emitters in the $^{223}$Ra decay chain. Qualitatively, the $^{215}$Po peak intensities show the same temperature dependence as $^{219}$Rn. However, they are difficult to interpret because, being the granddaughter of the $^{223}$Ra source, not all $^{215}$Po originate from the source, but from all those places where $^{219}$Rn has ended up, and because its half-life (1.8 ms) is comparable to the transport time. The origin of $^{211}$Bi recoil ions, being three alpha decays away from $^{223}$Ra, is even more uncertain than that of $^{215}$Po. Moreover, $^{211}$Bi is reached via $^{211}$Pb that has a half-life of 36 minutes, longer than our typical measurement time for a certain temperature setting. This prevents us to deduce any relevant information from the $^{211}$Bi peak intensities.

The experimental cell is attached to the so-called 1 K pot of the bath cryostat recently used in our development of a superfluid helium ion catcher [14,15]. Pumping the cryogenic liquid in the 1 K pot lowers its boiling point. With liquid helium we can cool down to 1.0 K. For the data shown here only liquid nitrogen was used to reach temperatures down to about 60 K. The temperature is kept constant by a LakeShore Model 340 temperature controller in combination with a Pt100 temperature sensor and a heater, both installed on the 1 K pot.

As the experiments reported here aimed to show the effect of the freezing out of impurities, we did not use high purity gas in the experimental cell at room temperature. Therefore, the experimental cell was not at all designed for good pumping capability. Moreover, in the case of helium, the impurity level at room temperature was intentionally increased by adding air to the cell before filling it with helium. If uniformly frozen on all surfaces, 3 mbar of air would form a solid layer on top of the source, thick enough to stop the recoils. To prevent this, we chose a 10 times lower partial air pressure (0.3 mbar) to be on the safe side. This corresponds to an impurity level of about 300 ppm or a density of $1.0 \times 10^{16}$ atoms/cm$^3$, consisting mostly of air. For neon and argon, no extra air was added, and we estimate the impurity content to be around 100 ppm.

At the start of a measurement series, the cell is filled at room temperature with 1 bar noble gas (and 0.3 mbar of air in the case of helium) as described above. The cell is then closed and cooled down to the lowest temperature, at constant gas density in the cell. For argon gas, measurements are restricted to above 75 K to avoid its condensation. The temperature is then increased in steps and the alpha particle energy spectrum measured. At each step, thermal equilibrium between cell and the noble gas is confirmed by consecutive measurements showing the same peak intensities. Measurements above about 200 K are not useful as radon is not frozen out and the $^{219}$Rn atoms are floating through the chamber (this is readily apparent from the alpha-energy spectra).

Figure 1 indicates the voltages applied to the source, the guiding electrode, and the aluminum foil for helium and argon in the cell. For neon these voltages were lowered to 65 % to prevent electrical discharging. This reduces the magnitude of the electric field without changing its profile . Thus for neon the ion tracks are the same as in Figure 1, only the ion velocity is reduced.

Figure 2 shows a typical alpha particle energy spectrum and its analysis. The efficiency that $^{219}$Rn recoil ions survive thermalization and transport to the aluminum foil in an ionic state, measured as a function of temperature for helium, neon and argon gas, is shown in Figure 3. Towards lower temperatures the efficiency is strongly enhanced, starting at about 120 K for all three noble gases. It saturates below about 90 K. For helium gas, in a separate experiment related to the study of ions in superfluid helium, the saturation efficiency did not change anymore down to 4 K. A least-squares fit of the data using a Hill's equation was performed. The saturation efficiency at low-temperatures determined as one of the fit parameters is 28.7(1) % in helium, 22.1(2) % in neon, and 17.0(2) % in argon. Indicated are only the statistical errors from the measured spectra and the fitting procedure. The overall systematic error of about 6 % is due to the uncertainty in determining the detector solid angle.

We believe the high efficiencies and their saturation at low temperatures to be due to the freezing out of impurities which enhances the survival probability of the thermalized ions. The saturation efficiency is reached once all impurities are frozen out. In our set-up, we cannot distinguish between atomic or molecular ions reaching the aluminum foil in front of the detector. At high temperatures, our measured efficiencies include both variants; although molecular ions should be rather unlikely since radon is a noble gas. At low temperatures, however, impurities that could take part in forming molecules are frozen out first, leaving the measured efficiency solely to the atomic $^{219}$Rn ions.

In the absence of impurities, the fate of the ions can still be affected by the ionization of the noble gas. We checked this qualitatively for helium: the electron-ion density was altered by changing the gas density and the electric field strength. A higher electric field separates electrons and ions quicker, resulting in a lower equilibrium ionization density, while a higher pressure increases the ionization density. We found that the efficiency at low temperatures (73 K) in the saturation regime does not change when the electric field is lowered by a factor of 4 or when the gas density is changed by a factor between 0.25 and 1.6. From this we conclude that at thermal energies neutralization due to free electrons is negligible in our experiments. Therefore the observed saturation efficiencies at low temperatures may well reflect the charge exchange and stripping cross sections involved in the slowing-down process of the ions. This would represent a fundamental upper limit for the efficiency of noble gas ion catcher devices. Note that the efficiencies decrease systematically when going from helium over neon to argon. This demonstrates that our results are based on intrinsic properties of the noble gases. A similar trend has been observed earlier for the slowing down of protons and muons in noble gases [16,17].

Our data have to be compared with those obtained for ultra-pure helium gas catchers at room temperature. During the development of such devices by G. Savard and collaborators [7,18,19], efficiencies of up to 45% for fission fragments from a $^{252}$Cf source were achieved. Taking into account the systematic uncertainties involved and the fact that some dependence on the specific element has to be expected, we consider those measurements compatible with our work.

Huikari et al. [11] used a room-temperature gas catcher system with modest purity helium gas (of order several ppm) to also extract $^{219}$Rn recoils from a $^{223}$Ra source. They reported a $^{219}$Rn efficiency of 75% at a helium pressure of 50 mbar, decreasing strongly with increasing pressure: down to 20% at the lowest density used in the present work (equivalent to 250 mbar at room temperature) and down to an extrapolated value of 0.2% at a density equivalent to 1 bar at room temperature. This latter value is only slightly lower than our detection limit of about 0.3% efficiency. However, the pressure dependence of the efficiency reported in [11] differs completely from that observed by us and also by Maier et al. [18] for ultra-pure helium, where the efficiency does not change at all in the investigated pressure range. One could speculate whether the strong decrease of efficiency with increasing pressure observed in [11] is due to the presence of impurities (e.g. due to an increased neutralization via three-body reactions with electrons and impurity molecules). The fact that efficiencies higher than our low-temperature saturation values are obtained (at pressures below about 200 mbar) would indicate a higher ion survival probability during slowing down in the presence of impurities. This would happen if charge exchange and stripping cross sections on impurity molecules are more favourable towards ion survival than those on helium atoms. The fact that in the system used by Huikari et al. [11] molecular ions are broken up in the extraction region of the gas catcher and contribute to the measured efficiency may be relevant in this discussion. Dedicated work is needed to understand this issue. Since gas catchers have to operate at a rather high pressure to provide enough stopping power, there is no question that ultra-pure conditions have to be employed.

We have demonstrated large stopping and transport efficiencies of ions in noble gas stopping media of low purity in a container that is not ultra-high vacuum compatible by in-situ purification of the noble gas upon cooling to below 90 K. The measured efficiencies at low temperature are comparable to those achieved with ultra-high purity gas catchers at room-temperature. However, we consider constructing a cryogenic noble gas ion catcher operating at liquid nitrogen temperature to be technically easier. It may therefore be a more practical choice.

**Acknowledgement**

We would like to thank H. Penttilä and S. Rinta-Antila from the Department of Physics of the University of Jyväskylä (Finland) for providing us with the $^{223}$Ra source.

**Figure captions.**

Figure 1.

Schematic view of the experimental cell. The voltages used when filled with helium and argon gas are indicated. For neon, voltages were reduced to 65 % of these values. The detector and container are grounded. Equipotential lines and ion trajectories resulting from ion optics simulations using SIMION [13] are shown; they indicate a 100 % ion transport efficiency. In the simulations, the presence of the gas is taken into account by a viscous force.

Figure 2.

Alpha particle energy spectrum, (400 s measurement time), and its analysis for helium gas at 108 K. The solid line at the top is the measured spectrum. The fitted spectrum is overlayed in gray/green. The individual components resulting from the fit for the three alpha decaying isotopes originating from both the foil (black/red) and the source (gray/blue) are shown in the lower parts. The fit takes into account the known energy and intensity relationships.

Figure 3.

Measured efficiency vs. temperature that 100 keV $^{219}$Rn ions survive thermalization and transportation over about 3 cm in an ionic state in 1 bar (at room temperature) noble gases. The lines are least-squares fits of a Hill's equation to the data. The saturation efficiency (with statistical error) deduced from the fit is indicated. The overall systematic uncertainty amounts to about 6 % as discussed in the text.

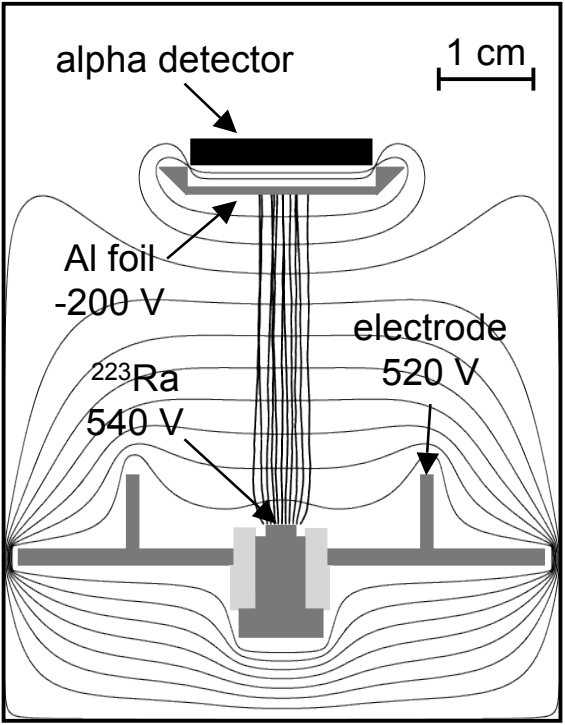

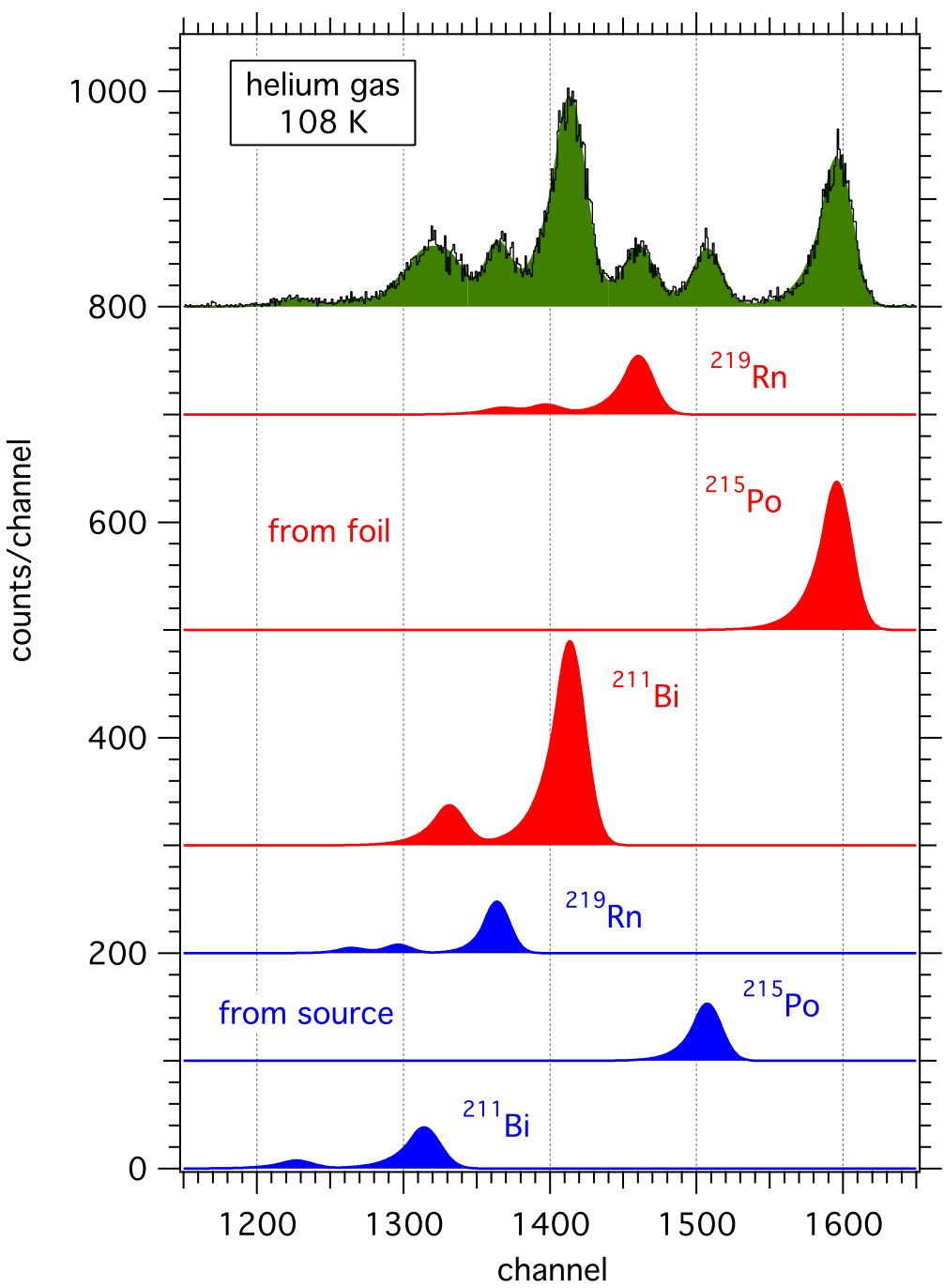

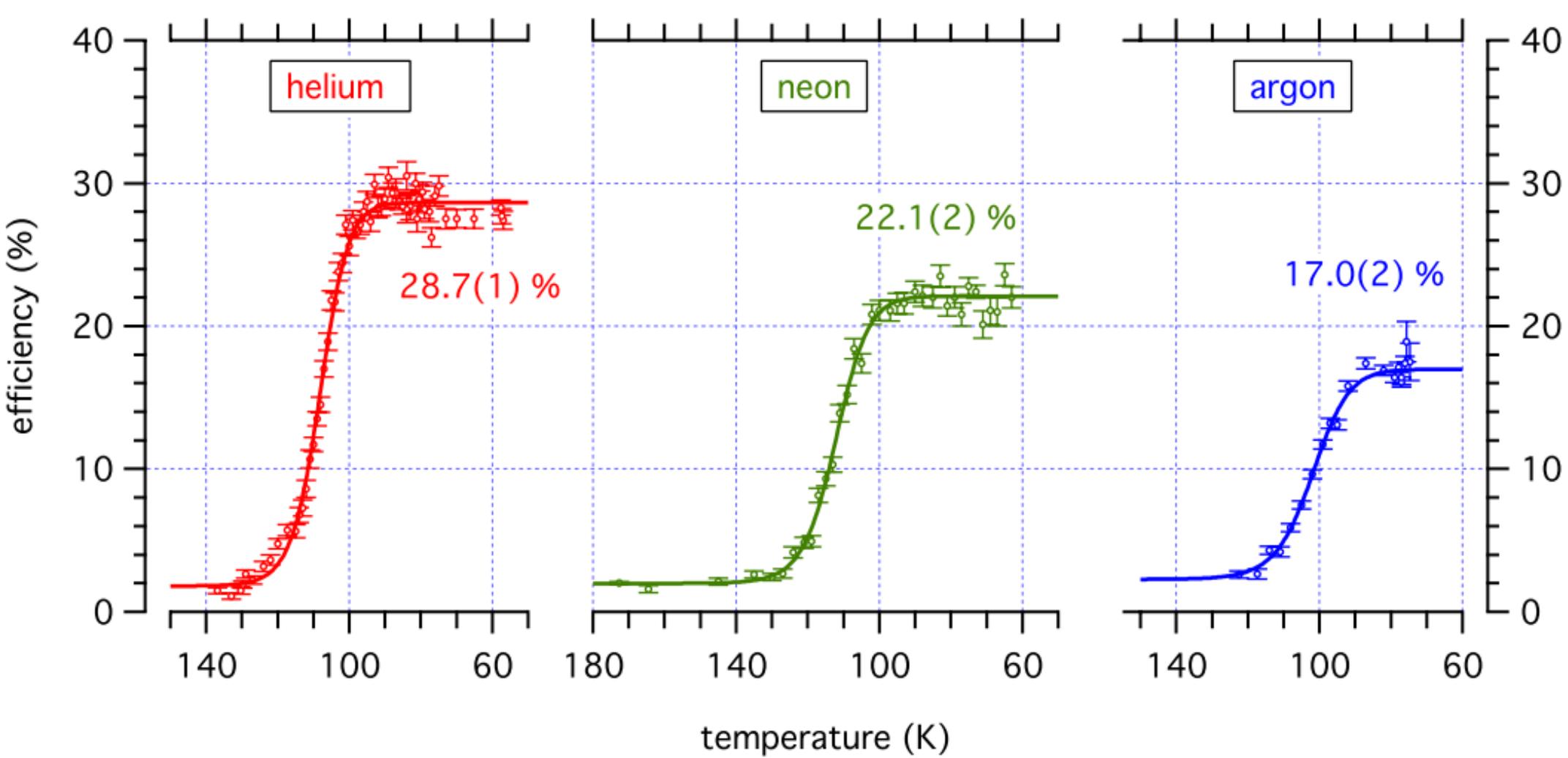